\documentclass[journal=jacsat,manuscript=article]{achemso}

\usepackage[version=3]{mhchem} 
\usepackage{gensymb}
\usepackage{xcolor}
\usepackage{bm}
\usepackage{xr-hyper}
\usepackage{hyperref}

\title{Molecular Fingerprints of Ice Surfaces in Sum Frequency Generation Spectra: a First Principles Machine Learning Study} 

\author{Margaret L. Berrens}
\affiliation[1]
{Department of Chemistry, University of California Davis, One Shields Ave. Davis, CA, 95616.}
\author{Marcos F. Calegari Andrade }
\affiliation{Quantum Simulations Group, Materials Science Division, Lawrence Livermore National Laboratory, Livermore, California, 94550-5507}
\alsoaffiliation{Laboratory for Energy Applications for the Future, Lawrence Livermore National Laboratory, Livermore, California, 94550-5507}
\author{John T. Fourkas }
\affiliation{Department of Chemistry and Biochemistry, University of Maryland, College Park, MD, 20742.}
\alsoaffiliation{Institute for Physical Science and Technology, University of Maryland, College Park, MD, 20742.}
\alsoaffiliation{Maryland Quantum Materials Center, University of Maryland, College Park, MD, 20742.}
\author{Tuan Anh Pham}
\email{pham16@llnl.gov}
\affiliation{Quantum Simulations Group, Materials Science Division, Lawrence Livermore National Laboratory, Livermore, California, 94550-5507}
\alsoaffiliation{Laboratory for Energy Applications for the Future, Lawrence Livermore National Laboratory, Livermore, California, 94550-5507}
\author{Davide Donadio}
\email{ddonadio@ucdavis.edu}
\affiliation[1]
{Department of Chemistry, University of California Davis, One Shields Ave. Davis, CA, 95616.}


\begin{document}







\begin{abstract}
Understanding the molecular-level structure and dynamics of ice surfaces is crucial for deciphering several chemical, physical, and atmospheric processes. Vibrational sum-frequency generation (SFG) spectroscopy is the most prominent tool for probing the molecular-level structure of the air--ice interface as it is a surface-specific technique, but the molecular interpretation of SFG spectra is challenging. This study utilizes a machine-learning potential, along with dipole and polarizability models trained on \textit{ab initio} data, to calculate the SFG spectrum of the air–ice interface. At temperatures below ice surface premelting, our simulations support the presence of a proton-ordered arrangement at the Ice I$_h$ surface, similar to that seen in Ice XI. Additionally, our simulations provide insight into the assignment of SFG peaks to specific molecular configurations where possible and assess the contribution of subsurface layers to the overall SFG spectrum. These insights enhance our understanding and interpretation of vibrational studies of environmental chemistry at the ice surface.

\end{abstract}



\section{Introduction}
The surface of ice provides a distinct molecular environment that catalyzes environmentally impactful chemical reactions.\cite{watanabe_ice_2008, park_fundamental_2010, delzeit_ice_1997} Terrestrial ice formations such as snowflakes, glaciers, and ice particles within cirrus and polar stratospheric clouds play a crucial role in shaping the atmosphere's chemical makeup and heat distribution.\cite{blanco-alegre_role_2024, barrie_atmospheric_1985, zondlo_chemistry_2000} The characteristics of ice surfaces undergo significant changes with temperature. At temperatures below $\sim$ 170 K, ice exhibits a crystalline surface, with water molecules frozen in position.\cite{wei_sum-frequency_2002, suter_surface_2006} As the temperature rises slightly, surface molecule mobility increases while the bulk molecules remain relatively static. Further temperature elevation to $\sim$ 200 K enhances surface molecule mobility, leading to surface premelting and the formation of a quasi-liquid layer.\cite{wei_sum-frequency_2002, wei_vib_2002, wei_sfg_2001} For ice surfaces below the premelting threshold, surface water molecules remain fixed over short observation periods. However, over longer timescales, surface molecules can migrate creating conditions conducive to reactions, albeit at a slower rate than in the quasi-liquid layer.\cite{park_fundamental_2010, lee_ammonia_2023, kim_segregation_2011} Therefore, understanding the structure, dynamics, and local hydrogen bonding environment of the ice surface is crucial for gaining insights into reactions occurring in conditions below the premelting range, as in space or stratospheric clouds. 

Specifically, proton ordering on the surface of ice at temperatures below 180 K has been suggested to affect the adsorption of polar monomers, impacting physical and chemical reactions in ice clouds and ice growth.\cite{sun_role_2012} Despite hexagonal ice having a proton-disordered bulk, its surface may exhibit proton ordering, but to what extent remains unknown. There is thermodynamic evidence for
an ordered striped surface\cite{buch_proton_2008, pan_surface_2008} However, in naturally occurring environments, ice surfaces may not be in the lowest free energy state. Further theoretical studies, such as \textit{ab initio} molecular dynamics (MD) simulations, and experiments, are necessary to determine the extent and nature of proton ordering to further determine its influence on the properties of ice surfaces, including their interaction with adsorbed species.

Experimentally probing the surface of ice and its proton ordering presents difficulties due to the vastly greater number of molecules in the bulk phase. It is essential to use a surface-sensitive and minimally invasive technique, as the ice surface is susceptible to melting and irreversible deformation.\cite{smit_observation_2017} 
Sum frequency generation (SFG) spectroscopy has proven to be a powerful technique for studying the surface of ice as this technique is noninvasive, highly surface-specific, and selective to molecular groups.\cite{morita_theory_2018, tang_molecular_2020} 
There are two ways of using SFG, the homodyne and the heterodyne mode. 
Homodyne SFG measures the magnitude of the second-order nonlinear susceptibility $|\chi^{(2)}|^2$,  while heterodyne SFG can measure both the imaginary part (Im$\chi^{(2)}$) and the real part (Re$\chi^{(2)}$) of the second-order nonlinear susceptibility. In homodyne SFG, $|\chi^{(2)}|^2$ provides a measure of the overall intensity of the SFG signal, reflecting the magnitude of the nonlinear response of the system but without detailed phase information. In heterodyne SFG, the Im$\chi^{(2)}$ spectrum reveals information about the orientation of the transition dipole moment, which provides insights into the molecular orientation for a given vibrational mode.\cite{sun_phase-sensitive_2019, mirzajani_accurate_2022} 
Whereas heterodyne SFG requires more careful interpretation and experimentation it can lead to a more comprehensive understanding of molecular orientations and interactions at interfaces.\cite{rivera_reexamining_2011, pool_comparative_2011} Being able to assign SFG peaks 
to specific molecular configurations of water would greatly enhance the interpretation of vibrational spectroscopy experiments at the air–ice interface, especially when adsorbed molecules are present. Additionally, determining contributions from subsurface layers would clarify how deeply adsorbed molecules penetrate the ice surface, further aiding in the interpretation of experimental SFG spectra.

There are limited published experimental SFG spectra of the air--ice interface at temperatures lower than the onset of premelting, and those available often show discrepancies. Notably, there is a controversy regarding the O--H stretch mode in the 3000 $\leq$ $\omega$ $\leq$ 3400 cm$^{–1}$ frequency range of heterodyne SFG spectra. 
Nojima \textit{et al.}\cite{nojima_proton_2017} reported large positive feature in the Im$\chi_{ssp}^{(2)}$ spectrum for ice I$_h$ at 130 K. They attribute this peak to the surface with “H-up” proton order, implying that a majority fraction of water molecules near the air--ice interface point their free O--H upward to the air, suggesting that proton disorder is not preserved at the surface of ice I$_h$.\cite{nojima_hydrogen_2020} 
In contrast, Smit \textit{et al.} \cite{smit_observation_2017} observed small positive at 3110 cm$^{–1}$  and large negative feature at 3150 cm$^{–1}$  in the Im$\chi_{ssp}^{(2)}$ spectrum at 150 K. They attribute these features to a bulk response through the electric quadrupole transition.\cite{smit_observation_2017, ishiyama_origin_2012, shultz_multiplexed_2010} Yamaguchi \textit{et al.}\cite{yamaguchi_perspective_2019-1} attribute these experimental discrepancies to differences in the experimental setup. Due to difficulties in carrying out SFG experiments at the air--ice interface, it is important to be cautious with the interpretation of the peak between 3000 $\leq$ $\omega$ $\leq$ 3400 cm$^{-1}$ in the O--H stretching region of the Im$\chi_{ssp}^{(2)}$ SFG spectrum. As for the O--H stretch mode in the 3400 $\leq$ $\omega$ $\leq$ 4000 cm$^{–1}$ frequency region, Smit \textit{et al.} \cite{smit_observation_2017} report a broad positive peak around 3530 cm$^{–1}$ and another around 3700 cm$^{–1}$, corresponding to the free O--H stretch at the water/ice surface. Currently, experimental estimates for the low-frequency region of the heterodyne SFG spectra of the ice surface are lacking, mainly due to measurement difficulties. 
Experimental SFG data alone are insufficient for correlating spectroscopic observations with molecular structure, making atomistic simulations necessary for a comprehensive microscopic understanding. 

Several theoretical studies have characterized the SFG spectrum of the air--ice interface, employing various system sizes, time scales, and theoretical approaches.  Smit \textit{et al.} \cite{smit_observation_2017} uses classical MD simulations to interpret only the positive feature in the experimental Im$\chi_{ssp}^{(2)}$ spectrum at 3530 cm$^{–1}$. They attribute the feature mainly to the asymmetric OH stretch of four coordinated molecules at the surface of ice. Ishiyama \textit{et al.}\cite{ishiyama_direct_2014} used QM/MM to calculate the heterodyne and homodyne SFG spectra of the ice surface at 130 K. Their heterodyne spectrum aligns with the experimental spectrum of Smit \textit{et al.}\cite{smit_observation_2017} in the 3000 $\leq$ $\omega$ $\leq$ 3400 cm$^{-1}$ range, which they attribute to asymmetric contributions of charge transfer from the interstitial bilayer region. However, their spectrum lacks the high-frequency positive peak in the 3400 $\leq$ $\omega$ $\leq$ 3600 cm$^{-1}$ range and the free O--H peak, suggesting potential issues with convergence. 
Wan and Galli\cite{wan_first-principles_2015} used perturbation theory to investigate the homodyne SFG spectrum. They incorporated bulk quadrupolar contributions to their calculations and found that these contributions produce appreciable differences in the 3000 $\leq$ $\omega$ $\leq$ 3400 cm$^{-1}$ range of the O--H stretching band, supporting the findings of experimental results.\cite{smit_observation_2017, ishiyama_origin_2012, shultz_multiplexed_2010}
A comprehensive review of previous experimental and theoretical work on the SFG spectrum of the ice surface highlights key findings and discrepancies.\cite{tang_molecular_2020} Despite the range of reported experimental and theoretical SFG spectra, a unified understanding of the SFG spectrum at the air--ice interface is still lacking.

\begin{figure}[h]
    \centering
    \includegraphics[width=0.7\linewidth]{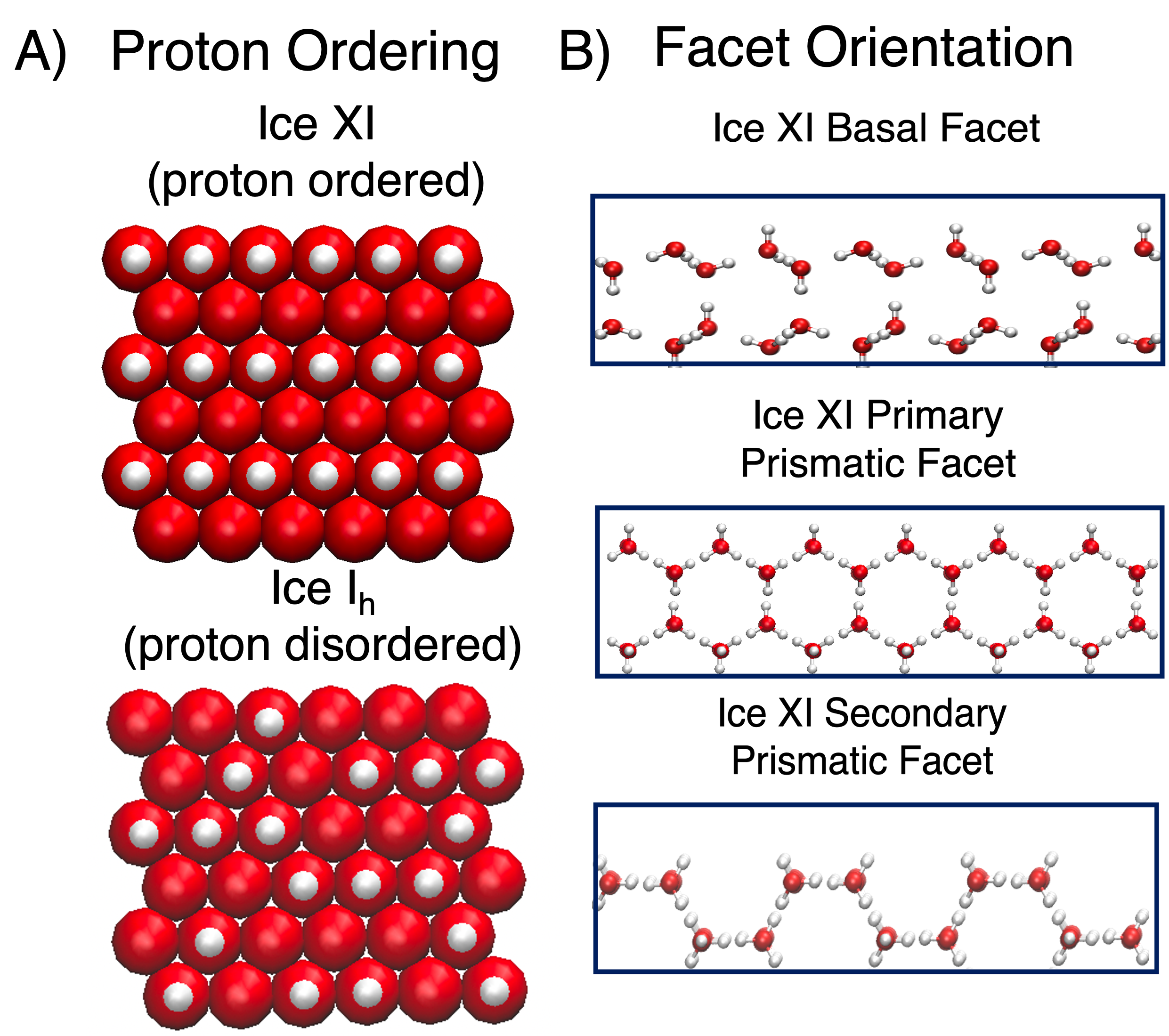}
    \caption{A) Top-down snapshots of the proton arrangement of the free O--H bond at the surface of Ice XI (striped phase) and Ice I$_h$ (one example of a proton ordered configuration). B) Side snapshots of the structure of top layers from the surface the three low-index facets sampled for Ice XI (basal, primary prismatic, and secondary prismatic). }
    \label{fig:icestructure}
\end{figure}

To obtain a comprehensive understanding of the SFG spectrum at the air--ice interface and the proton ordering on the I$_h$ surface, we perform a detailed characterization of the vibrational spectrum of the ice surface using classical MD simulations. This is accomplished by utilizing machine-learned potentials (MLPs), dipole, and polarizability models trained on \textit{ab initio} data, enabling large-scale and long-time MD simulations for on-the-fly SFG spectrum calculations. It is known that calculating the SFG spectrum with MD simulations has a slow convergence.\cite{morita_theoretical_2002, moberg_temperature_2018} For this reason, we harness the speed-up of at least three orders of magnitude, relative to first principles MD simulations, that can be gained by using machine-learned models.\cite{litman_fully_2023} Here, we focus on examining subtle structural differences in the vibrational spectrum, specifically investigating proton ordering at the ice surface and facet orientation. Figure~\ref{fig:icestructure}a) shows representative top-down snapshots of the proton ordered (striped phase\cite{buch_proton_2008}) arrangement of the free O--H bonds seen at the surface of Ice XI, as well as a representative proton configuration of the free O--H bonds at the surface of Ice I$_h$. Figure~\ref{fig:icestructure}b) shows side snapshots of the surface structures of the three low-index facets sampled. This work helps determine whether the surface of ice below 180 K exhibits a proton-ordered arrangement. Additionally, we address the assignment of SFG peaks to specific molecular configurations where possible and evaluate the contribution of subsurface layers to the overall SFG spectrum. By investigating these aspects, we provide a deeper understanding of the molecular-level interactions and structural dynamics at the ice surface, thereby shedding light on the complex mechanisms governing ice surface chemistry. 

\section{Results and Discussion}

\subsection{ Ice XI vs. Ice I$_h$}

\begin{figure}[h!]
    \centering
    \includegraphics[width=1\linewidth]{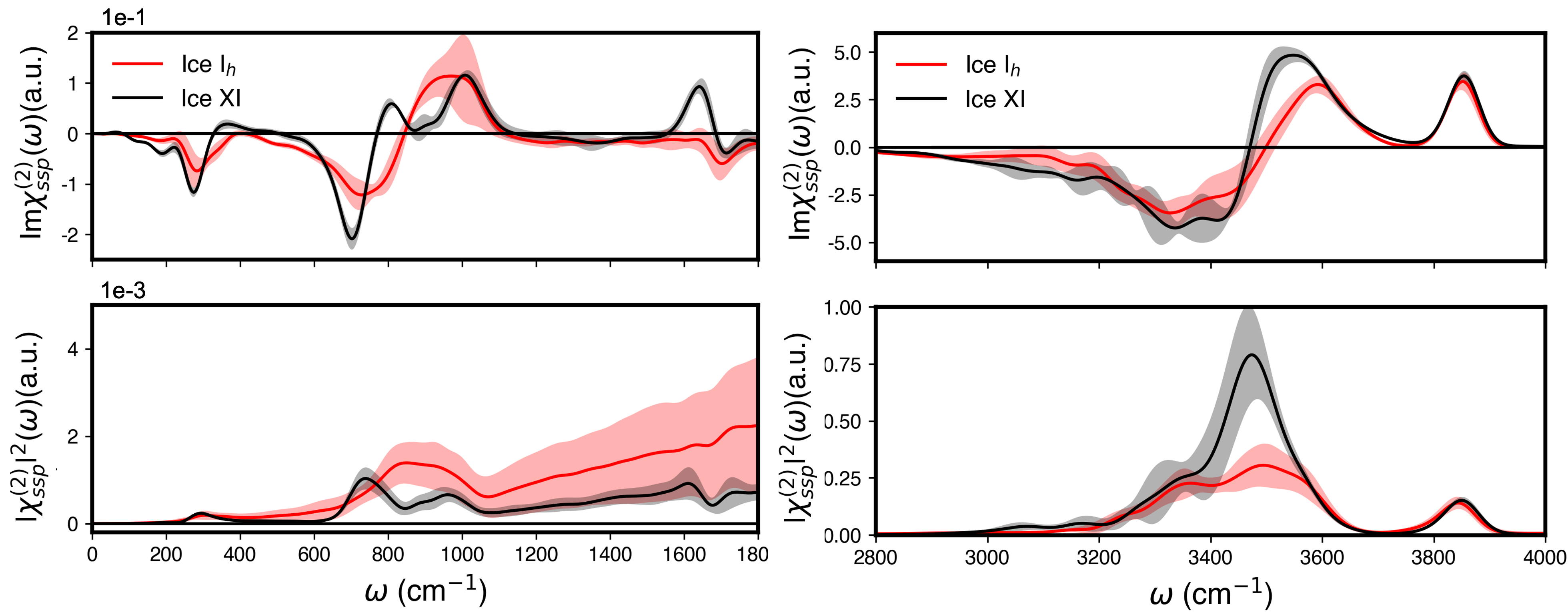}
    \caption{Homodyne (bottom panels) and heterodyne (top panels) SFG spectra for antiferroelectric Ice XI and Ice I$_h$ for the low-frequency (left) and high-frequency regions (right). Ice XI spectra are averaged over four runs, Ice I$_h$ spectra are averaged over nine configurations, and each simulation ran for 4~ns. Shaded regions represent the standard deviations over the different runs.}
    \label{fig:protonorder}
\end{figure}

Figure~\ref{fig:protonorder} displays the theoretically calculated SFG spectra with $ssp$ polarization in the high (right panels) and low-frequency (left panels) regions of the vibrational spectra of the basal facet of both Ice XI and Ice I$_h$. 
The low-frequency spectra (left panels) feature peaks between 1600 and 1800~cm$^{-1}$ corresponding to the molecular bending modes, pronounced broader bands between 400 and 1100~cm$^{-1}$ corresponding to librations, and weaker peaks below 400~cm$^{-1}$ corresponding to rigid translational modes.\cite{chen_role_2008}
In the high-frequency region (right panels), the range of 2800 - 3450 cm$^{-1}$ corresponds to the strongly hydrogen-bonded O--H stretch region, the range of 3450 - 3700 cm$^{-1}$ corresponds to the weakly hydrogen-bonded O--H stretch region, and the peak at around 3850 cm$^{-1}$ corresponds to the free O--H peak present at the surface of ice.\cite{smit_excess_2017} 
As we are not aware of published experimental SFG spectra probing the low-frequency region for ice surfaces, we can compare only the high-frequency region to experiments. 
We note that the O--H stretching frequencies blue-shifted compared to experiments in part because classical MD simulations do not account for nuclear quantum effects, which produce a red-shift in the free O--H peak. \cite{ohto_accessing_2019-2} 
Aside from this shift, we find that the overall shape of both of our Im$\chi^{(2)}$ and $|\chi^{(2)}|^2$ SFG spectra are in good agreement with the experimental spectra from Smit \textit{et al.}.\cite{smit_observation_2017} Notably, this includes the presence of the free O–H peak, the relatively sharp positive peak in the weakly hydrogen-bonded O--H stretching region, and the broad negative peak in the strongly hydrogen-bonded O--H stretching region of the Im$\chi^{(2)}$ spectra. As well as the positive peak in the weakly hydrogen-bonded O--H stretching region and relatively small positive free O--H peak in the $|\chi^{(2)}|^2$ spectra. Contrary to the measurements in Refs.~\cite{nojima_proton_2017, nojima_hydrogen_2020}, our calculated heterodyne spectrum does not show a positive peak in the 2800 - 3450 cm$^{-1}$. In the corresponding region of our spectra, we observe a negative amplitude absorptive feature in the Im$\chi^{(2)}$ spectrum and a corresponding dispersive feature in the Re$\chi^{(2)}$ spectrum (Figure~\ref{fig:full_spectrum}) in good agreement with previous calculations.\cite{wan_first-principles_2015, buch_sum_2007}
While our calculations may contribute to the discussion over the features observed experimentally in the 2800 - 3200 cm$^{-1}$ O--H stretching region, it cannot provide a definitive answer: this peak could be sensitive to small phase shifts in the experimental setup,\cite{yamaguchi_perspective_2019} but it may also originate from a bulk quadrupolar contribution that our simulations cannot capture.\cite{shultz_multiplexed_2010, wan_first-principles_2015}   


Both the homodyne and heterodyne SFG spectra exhibit significant differences in terms of peak positions and line shapes between proton-ordered ice XI and proton-disordered ice I$_h$ especially in the weakly hydrogen-bonded region and at low frequencies. 
The low-frequency spectrum of the basal plane of Ice XI is more structured than Ice I$_h$. In particular, the heterodyne spectrum for the bending mode of Ice XI has a positive (dispersive) and a negative (absorptive) peak at distinct frequencies but only an absorptive feature for the proton-disordered surface of Ice I$_h$.     
As for other water interfaces, this marked difference in the bending peak may be exploited to identify the ordering of the air--ice interface.\cite{seki_bending_2020, sudera_interfacial_2021}
The 600--1000 cm$^{-1}$ region of the SFG spectrum of Ice XI presents sharper features than that of Ice I$_h$. 

%
Our simulations demonstrate that SFG can differentiate between proton-ordered and proton-disordered surfaces in both the low and high-frequency regions of the vibrational spectrum, for both Im$\chi^{(2)}$ and $|\chi^{(2)}|^2$. Moreover, the proton-ordered $|\chi^{(2)}|^2$ SFG spectrum in the O--H stretching region aligns well with experimentally measured $|\chi^{(2)}|^2$ spectra of the air--ice interface at 150 K.\cite{smit_observation_2017, wei_surface_2001, sanchez_experimental_2017} It shows a small positive free O--H peak at 3850 cm$^{-1}$ along with a sharper more intense peak from 3450 - 3700 cm$^{-1}$ with a small tail from 2800 - 3450 cm$^{-1}$. In contrast, the $|\chi^{(2)}|^2$ spectrum for Ice I$_h$ has a rounded, less intense peak at 3450 - 3700 cm$^{-1}$, with a smaller intensity difference between it and the free O--H peak compared to Ice XI. This assertion is also supported by the Im$\chi^{(2)}$ spectra. The slightly larger intensity of the peak in the 3450 - 3700 cm$^{-1}$ region compared the free O--H peak for the Ice XI Im$\chi^{(2)}$ spectrum is reflected in Smit's  experimentally determined Im$\chi^{(2)}$ spectrum, unlike the equivalent peak intensities seen in our Ice I$_h$ spectrum. This finding supports previous work\cite{buch_proton_2008, pan_surface_2008, wan_first-principles_2015,hong_imaging_2024} suggesting that, at temperatures below the onset of disorder, the surface of ice has a proton-ordered arrangement similar to that of Ice XI. 


\begin{figure}[h]
    \centering
    \includegraphics[width=1\linewidth]{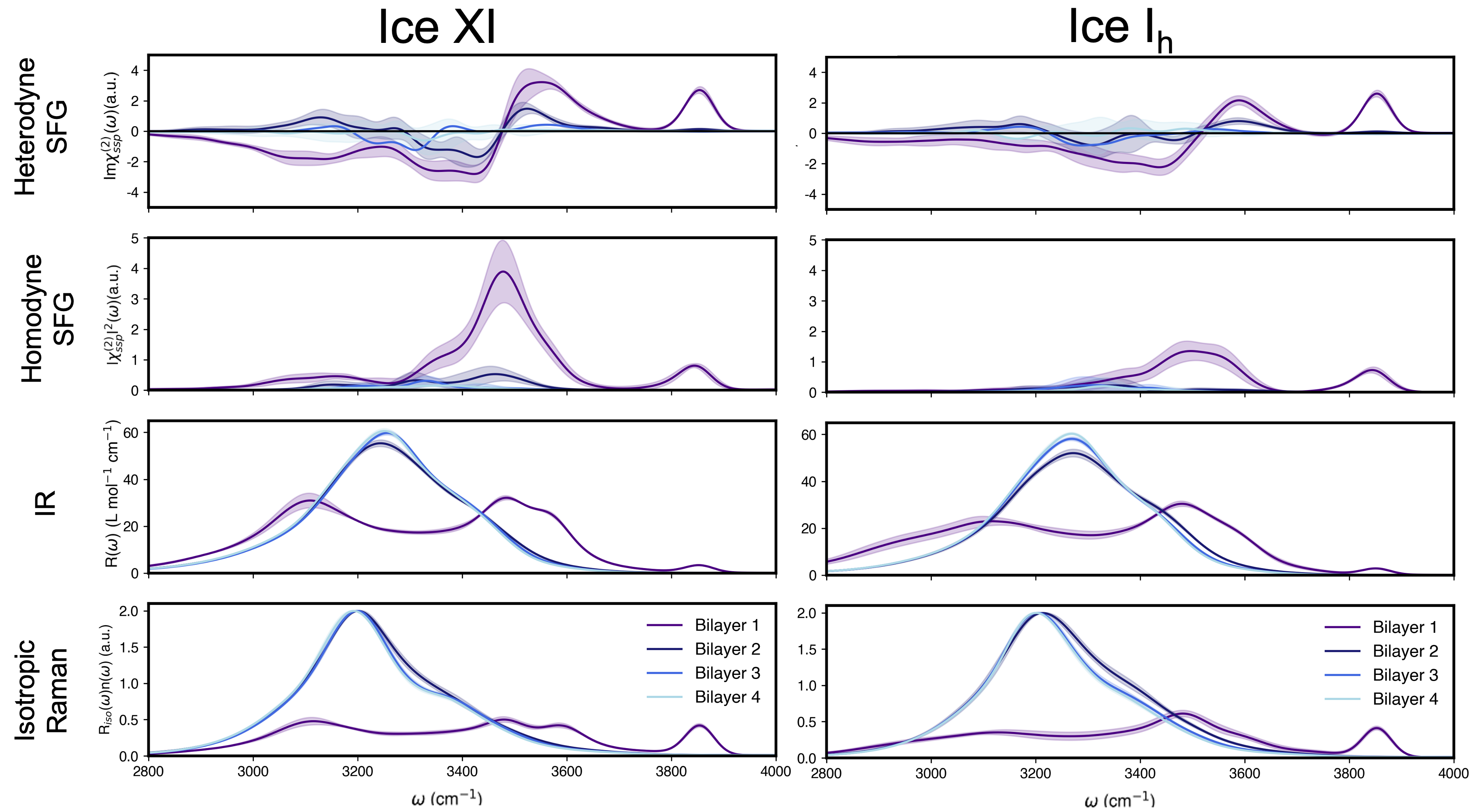}
    \caption{Heterodyne SFG, homodyne SFG, IR, and Raman spectra of the four bilayers from the surface for Antiferroelectric Ice XI and Ice I$_h$ for the O--H stretch region. Ice XI averaged over four configurations, Ice I$_h$ spectra averaged over nine configurations, each simulation ran for 2 ns. Shaded regions represent the standard deviations over the configurations. For the isotropic Raman spectra, $n(\omega) = 1 - e^{-\beta \hbar \omega}$ (the Bose Einstein factor).}
    \label{fig:bilayer}
\end{figure}

\subsection{Depth of SFG signal}

While some experimental techniques can be used to determine depth-resolved 
vibrational spectroscopy,\cite{balos_phase-sensitive_2022, fellows_obtaining_2023} the depth of the contribution to SFG has not been determined for the air--ice interface. Figure~\ref{fig:bilayer} represents the bilayer-resolved analysis of the Im$\chi^{(2)}$, $|\chi^{(2)}|^2$ SFG, IR, and isotropic Raman spectra for both Ice XI and Ice I$_h$ surfaces. The bilayer-by-bilayer anisotropic Raman spectra are reported in Figure~\ref{fig:bilayerir} and \ref{fig:icexi_raman_ir} for the top four bilayers from the surface. 
The results indicate that the top bilayer dominates the SFG spectra for both ice phases. 
Notably, the second bilayer also contributes to the spectra, with this effect being more pronounced in Ice XI than in Ice I$_h$. In contrast, the third and fourth bilayers have negligible contributions to the Im$\chi^{(2)}$ and $|\chi^{(2)}|^2$. The second bilayer contributes due to the absence of centrosymmetry in the region between the bilayers, resulting from its interactions with the first layer, as discussed in Figure~\ref{fig:bilayer_contrib_SI}. 
This is in agreement with the QM/MM calculations from Ishiyama \textit{et al.}\cite{ishiyama_direct_2014}. 
Comparing SFG to the IR and Raman spectra for both ice structures, we can see that there is a strong bulk IR response at 3250 cm$^{-1}$ and a strong bulk isotropic Raman response at 3200 cm$^{-1}$ from bilayers 2-4, that corresponds to the small contributions to the SFG spectra. 
The positive peak observed experimentally in the SFG within the 2800 - 3200 cm$^{-1}$ region aligns with the bulk IR and Raman response of ice seen here, in favor of the suggestion that the response could be a bulk quadrupolar contribution that is not captured by our simulated SFG under the dipolar approximation.\cite{shultz_multiplexed_2010, wan_first-principles_2015, smit_observation_2017, ishiyama_origin_2012, smit_supercooled_2017} 

Additionally, the differences seen in bilayer 1 from 3450 - 3700 cm$^{-1}$ of the SFG between Ice XI and Ice I$_h$ are also seen when comparing the IR and isotropic Raman spectra between the two structures. Specifically, the Ice XI spectra have two peaks in this region with a slight increase in the intensity of the shoulder compared to Ice I$_h$ at 3600 cm$^{-1}$. This analysis suggests that the SFG spectra of crystalline ice surfaces, regardless of whether these surfaces are proton-ordered or disordered, are primarily influenced by the top two bilayers. This result aids in the interpretation of experiments by indicating that vibrational spectroscopy of the ice surface, even at higher temperatures, primarily reflects the characteristics of these surface layers, thereby providing a clearer understanding of molecular interactions and surface properties.


\begin{figure}
    \centering
    \includegraphics[width=0.6\linewidth]{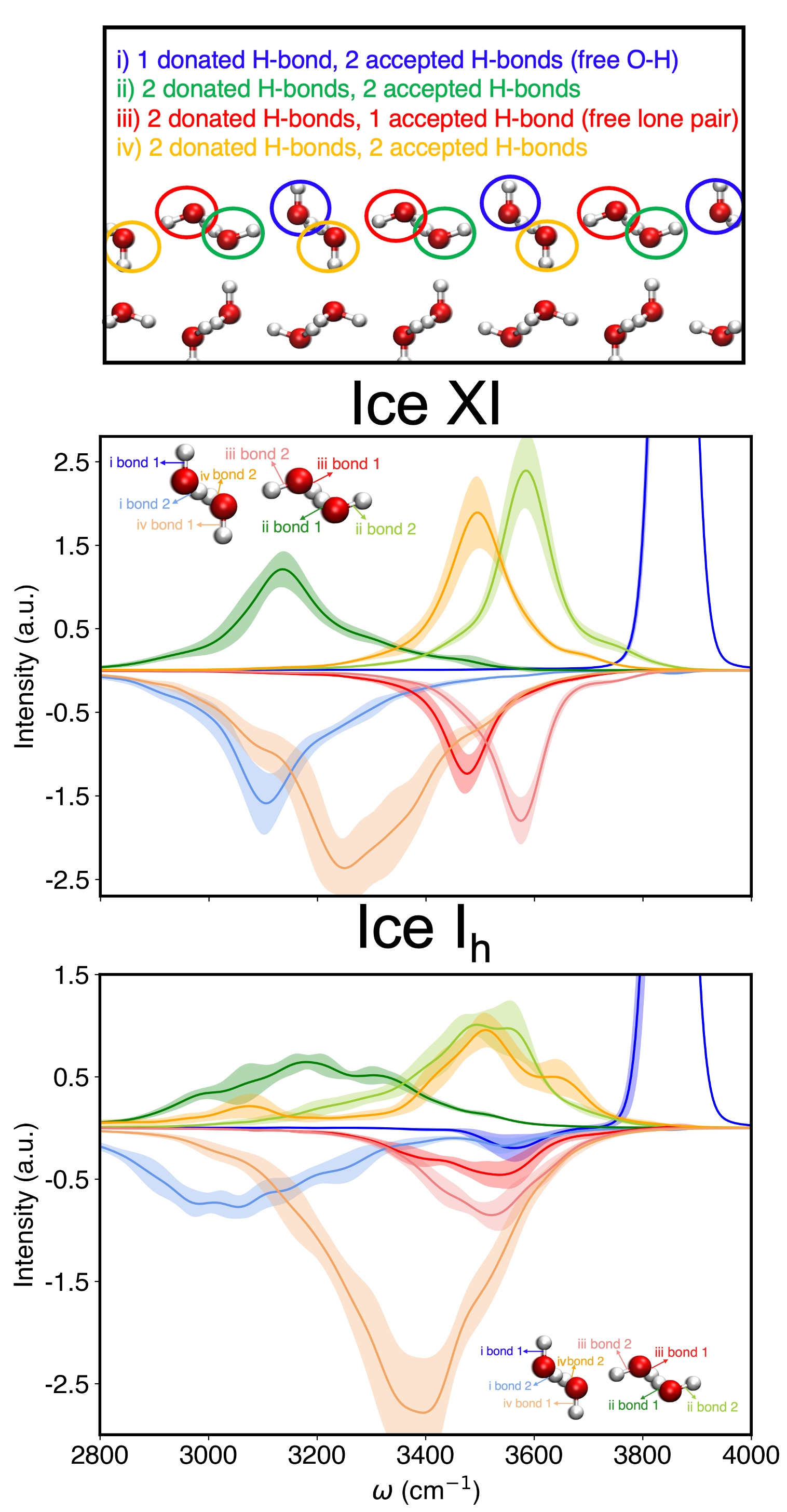}
    \caption{Orientation-weighted VDOS of individual O--H bonds of the four different water molecule orientations at the surface of Ice XI and Ice I$_h$ in the O--H stretching region of the vibrational spectra. Shown are the four types of water molecule orientations (top) the orientation-weighted VDOS assignment of the O--H stretching region of Ice XI (middle) and Ice I$_h$ (bottom). Shaded regions represent the standard deviation over blocks of 30 ps over the total 300 ps.}
    \label{fig:vdosassignment}
\end{figure}

\subsection{Molecular Peak Assignments}

Molecular simulations enable the calculations of the contributions of specific water molecules with given orientations to the overall vibrational spectrum. 
The basal plane of hexagonal ice features four possible orientations of water molecules with eight unique hydrogen-bond orientations (Figure~\ref{fig:vdosassignment}). In Ice I$_h$, the hydrogen bonding orientations of the molecules vary, whereas Ice XI has an ordered bonding pattern. 
To compare to heterodyne SFG spectra, we use the following orientation-weighted vibrational density of states (ow-VDOS)\cite{ohto_toward_2015} that captures the phasing of interfacial oscillators regardless of their SFG activity:
\begin{equation}
R_{ow-VDOS, (i,j)}(\omega) \propto \int_{0}^{\infty} dte^{-i\omega t} \left\langle {v_{z,(i,j)}(0)\frac{\bm{v}_{(i,j)}(t) \cdot \bm{r}_{(i,j)}(t)}{|\bm{r}_{(i,j)}(t)|}}\right\rangle
\end{equation}
where $(i,j)$ represents the molecule orientation type index $i$ (\romannumeral 1 -\romannumeral 4) with O--H bond $j$ (1-2), $\bm{v}_{z,(i,j)}(0)$ is the $z$ component of the velocity vector for $(i,j)$ at time 0, $\bm{v}_{(i,j)}(t)$ is the velocity vector at time $t$, and $\bm{r}_{(i,j)}(t)$ is the position vector for $(i,j)$ at time $t$. 
This equation provides orientational information to the vibrational spectra that are relevant to assigning features in Im$\chi^{(2)}$.\cite{superfine_experimental_1990, khatib_2016} 
Figure~\ref{fig:vdosassignment} shows the ow-VDOS for the O--H stretching region of the vibrational spectrum for Ice XI and Ice I$_h$ for the 8 unique O--H bond orientations at the surface of the ice slabs. 
The Ice XI ow-VDOS spectrum shows how the eight O--H bond types split into strong and weak hydrogen bonds, reflected in the shift of some spectra to higher frequencies and some shift toward lower frequencies, as seen previously.\cite{pham_first-principles_2012} This shift into lower and higher frequencies is also reflected in the reported bilayer by bilayer analysis of the Raman and IR spectra for the two structures as can be seen in Figure~\ref{fig:bilayer}. As reported in Table~\ref{tbl:hbondlengths}, these shifts are correlated with the differences in the hydrogen bond lengths of each O--H bond type.

While the overall intensity of peaks in SFG spectra may vary due to electronic effects, this orientation-weighted VDOS allows us to identify which types of bonds contribute to the different features of the OH stretching band. The peak at the highest frequency of ~3850 cm$^{-1}$ comes exclusively from (\romannumeral 1, 1), which is the free O--H stretch. Then in the weakly hydrogen-bonded region of 3450 - 3700 cm$^{-1}$ there are O--H stretching responses from four O--H bonds that fall into this region ((\romannumeral 2, 2), (\romannumeral 3, 1), (\romannumeral 3, 2), (\romannumeral 4, 2)). This region corresponds to the overall positive peak seen in the Im$\chi^{(2)}$ spectra, suggesting that there are stronger SFG intensities of the (\romannumeral 2, 2) and (\romannumeral 4, 2) O--H stretching responses compared to (\romannumeral 3, 1), (\romannumeral 3, 2). The O--H stretching responses from three O--H bonds falls into the 2800 - 3450 cm$^{-1}$ strongly hydrogen-bonded region ((\romannumeral 1, 2), (\romannumeral 2, 1), (\romannumeral 4, 1)). Here the orientation-weighted VDOS indicates the presence of a peak for (\romannumeral 4, 1) at 3400 cm$^{-1}$ (similar to bulk water) and for (\romannumeral 1, 2) at 3200 cm$^{-1}$ (similar to bulk ice). The O--H bond (\romannumeral 4, 1) points down towards the 2$^{nd}$ bilayer, and significantly contributes towards the negative absorptive peak at ~3400 cm$^{-1}$ in the overall SFG spectrum of the ice surface. The overall negative peak in the 2800 - 3450 cm$^{-1}$ region in our Im$\chi^{(2)}$ spectra indicates the strong intensities of the O--H stretching responses of (\romannumeral 1, 2) and (\romannumeral 4, 1) with respect to (\romannumeral 2, 1). 

In the orientation-weighted VDOS for Ice I$_h$, the relative frequency shifts of the peaks correspond to those of the Ice XI structure. However, each peak exhibits a broader shape and generally lower intensities, which is expected due to the disordered arrangement of protons. Notably, this results in a broader, lower intensity peak in Ice I$_h$ for (\romannumeral 1, 2), which is shifted towards lower frequencies (indicative of stronger hydrogen bonds). Simultaneously, the (\romannumeral 4, 1) peak is shifted to higher frequencies (indicative of weaker hydrogen bonds), exhibiting a broader shape and higher intensity. This effect is reflected in the overall higher intensity of the Ice XI, especially in the strongly hydrogen-bonded region seen in Figure~\ref{fig:protonorder}. The relative peak assignments as well as the depth analysis for both proton-ordered and proton-disordered surfaces offer valuable insights for future experimental and theoretical studies of adsorbed species on ice surfaces. This information can aid in the determination of whether these species preferentially interact with specific molecular orientations or molecules at certain depths from the surface, based on the peaks most altered in the SFG spectrum.\cite{seki_ions_2023}


\subsection{Other Facet Orientations}

\begin{figure}[h!]
    \centering
    \includegraphics[width=0.9\linewidth]{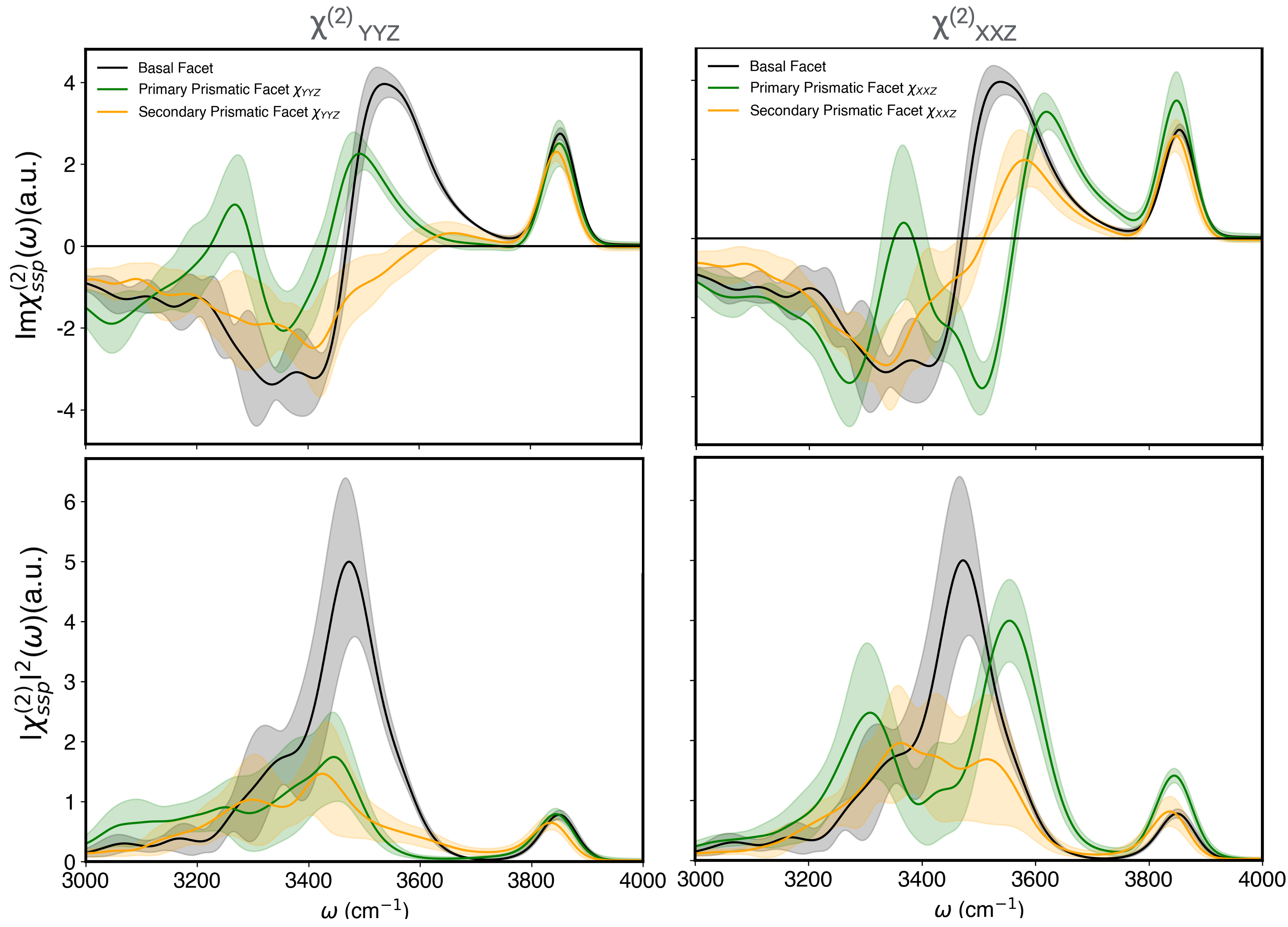}
    \caption{Homodyne (top) and heterodyne (bottom) SFG spectra for three low-index facets of antiferroelectric Ice XI. Each Ice XI facet averaged over four runs and each simulation ran for 4 ns. Shaded regions represent the standard deviations over the configurations.}
    \label{fig:facet}
\end{figure}

Crystal orientation provides valuable insights into the structure and bonding of ice surfaces. However, there are only a few experimental reports of the SFG spectrum for the three low-index facets of Ice XI and Ice I$_h$ (basal, primary prismatic, and secondary prismatic). Bisson \textit{et al.}\cite{bisson_hydrogen_2013}  examined Im$\chi^{(2)}$ for the secondary prismatic and basal facets at 100 K, identifying subtle differences, ascribed to different bonding motifs, emerging in the $|\chi^{(2)}_{ssp}|^2$ and $|\chi^{(2)}_{ppp}|^2$ spectra. Sanchez \textit{et al.}\cite{sanchez_experimental_2017} reported the $|\chi^{(2)}|^2$ spectrum experimentally for the basal and secondary prismatic facets at temperatures of 235~K and above, noting that the spectrum of the secondary prismatic facet appeared more rounded compared to the basal plane. 

Figure~\ref{fig:facet} presents our theoretical Im$\chi^{(2)}$ and $|\chi^{(2)}|^2$ spectra of the proton-ordered Ice XI surface in two polarizations. The basal facet has 3-fold rotational symmetry while the prism faces have only 2-fold rotational symmetry, necessitating spectra in different polarization combinations.\cite{bisson_hydrogen_2013} For all four reported spectra (Im$\chi^{(2)}_{xxz}$, Im$\chi^{(2)}_{yyz}$, $|\chi^{(2)}_{xxz}|^2$, $|\chi^{(2)}_{yyz}|^2$), we find that SFG can distinguish among the three low-index facets. As seen in Figure~\ref{fig:protonorder}, the Im$\chi^{(2)}$ and $|\chi^{(2)}|^2$ spectra for both polarizations are distinguishable in the high-frequency region of the O--H stretching spectrum. For the overall O--H stretching region, the primary prismatic facet exhibits four significant absorptive peaks, compared to three for the basal facet. Notably, in the low-frequency region of the O--H stretching spectrum, the negative peak in the primary prismatic spectra splits into two distinct peaks. Figure~\ref{fig:prim_vdos_assign} shows the VDOS assignment of the individual O--H bond types, from these assignments we can see the Y contributions tend towards lower frequency (stronger hydrogen bonds) while the X contributions tend towards higher frequency (weaker hydrogen bonds). The secondary prismatic facet has two absorptive peaks Im$\chi^{(2)}_{xxz}$ and three for for Im$\chi^{(2)}_{yyz}$. Specifically the secondary prismatic's Im$\chi^{(2)}_{xxz}$ as well as the $|\chi^{(2)}_{yyz}|^2$ resembles the Im$\chi^{(2)}_{ssp}$ and $|\chi^{(2)}_{ssp}|^2$ of the basal facet of Ice I$_h$. The origin of these similarities can be seen in Figure~\ref{fig:facet_structure} and \ref{fig:densityprofielsice}, where even after 1 ps the surface of the secondary prismatic facet reconstructs and becomes overall disordered, thus resembling the spectra of Ice I$_h$. The rounding of the secondary prismatic $|\chi^{(2)}|^2$'s peak from 3450 - 3700 cm$^{-1}$ region with respect to the sharp and intense peak in the basal facet has also been previously seen experimentally.\cite{sanchez_experimental_2017}

\section{Conclusions}
Our findings reveal that SFG is a sensitive probe capable of detecting subtle changes in the local hydrogen bonding environment, effectively differentiating between proton-ordered and disordered surfaces and their facet orientations. This work serves as the first reported SFG spectrum of the librational modes region of the air--ice interface, with appreciable differences between proton orderings in this region. Our calculations support the evidence of proton order at the surface of Ice I$_h$ at temperatures below the onset of premelting but above the transition temperature from Ice XI to Ice I$_h$. The surface bilayer is identified as the primary contributor to the SFG spectrum, with minor contributions from the second bilayer beneath the surface. Using the ow-VDOS spectrum, we pinpoint the individual O--H bond stretching contributions to the resulting SFG peaks. 
Finally, a comparison of the basal, primary prismatic, and secondary prismatic facets reveals that SFG spectroscopy can be used to probe the subtle structural differences between the facets. Overall, our comprehensive analysis of the vibrational spectrum of the air--ice interface provides an understanding of the sensitivity of SFG to small differences in molecular structure of the air--ice interface, that will aid in the interpretation of future vibrational studies of environmental chemistry at the ice surface. This study not only advances our fundamental knowledge of ice surfaces but also has broader implications for interpreting experiments of atmospheric chemistry, where ice plays a pivotal role. 

\section{Computational Methods}
We used an MLP to accelerate first principles simulations. We employed a previously trained and validated deep neural network potential (DNNP)\cite{zhang_phase_2021} that reproduces accurately structural, and vibrational\cite{msommers_raman_2020}  properties of water and ice, which we then augmented with long-range electrostatic interactions.\cite{zhang_why_2023} We ran classical MD simulations of an 864 molecule slab of antiferroelectric Ice XI and Ice I$_h$ systems (basal, primary prismatic, and secondary prismatic), each with 12 bilayers, at 211 K (100 degrees below the respective melting temperature of the MLP used).\cite{piaggi_phase_2021} Side snapshots of the slabs for Ice XI and Ice I$_h$ can be seen in Figure~\ref{fig:side_view_snap}. Each ice slab was generated using GenIce,\cite{matsumoto_genice_2018} which ensures the generation of completely randomized hydrogen-disordered networks obeying the ice rules for I$_h$ and ensures zero net polarization for both the Ice I$_h$ and Ice XI slabs. Convergence of the vibrational spectra on system size (number of bilayers) was checked (Figure~\ref{fig:size_converge}). Both systems had periodic boundary conditions (PBC) along the X and Y dimensions, with 70 \AA{} of vacuum region in the Z direction. The equations of motion were integrated with the velocity Verlet algorithm with a time step of 0.5 fs and the temperature was controlled by stochastic velocity rescaling\cite{bussi_canonical_2007} with a relaxation time of 1 ps. All simulations were performed using the DEEPMD-KIT\cite{wang_deepmd-kit_2018, zeng_deepmd-kit_2023} plugin with the LAMMPS package.\cite{thompson_lammps_2022}  Each system was equilibrated for 100 ps followed by respective production runs of 2-4 ns for SFG spectra, IR, and Raman spectra and 300 ps for vibrational density of states.

\subsection*{Training of Long Range Machine-Learned Interatomic Potential and Dipole/Polarizability Model}
An MLP for the air--ice interface was constructed using the long-ranged version of the Deep Potential method~\cite{zhang_deep_2022}. The machine-learning model is a 3 layers deep feedforward neural network (DNN) with 120 neurons per layer. The inputs to the DNN consist of the chemical environment within a $6$~\AA{} cutoff from each atom, smoothly decaying from 3~\AA{}. The data used to train this model were collected by reinforcement learning, in which water, hexagonal ice, cubic ice and their interfaces with air were explored at ambient pressure and temperatures ranging from 300 to 400 K using MLP-MD. Within this method, the MLP dataset was augmented at each iteration with atomic configurations exceeding a force error threshold of $0.1$ eV/\AA{}  re-evaluated with DFT. The force error metric used in reinforcement learning is the maximum deviation in atomic forces predicted by 3 independently trained MLP models. The atomic configurations in the MLP dataset also composed the dataset of two separate DNN models predicting the dipole and the static polarizability of water. These two models were independently optimized and their construction followed previously reported methods using DNN architectures similar to the Deep Potential model~\cite{msommers_raman_2020,zhang_deep_2020}.

\subsection*{Numerical Modeling of SFG Spectra}
Here, under the electric dipole approximation, the vibrationally resonant component of the second-order susceptibility $\chi_{pqr}^{(2)}$ is calculated as:\cite{morita_theoretical_2002, moberg_temperature_2018,litman_fully_2023}
\begin{equation}
\chi_{pqr}^{(2),R}(\omega_{IR}) = n_{BE}(\omega)\frac{i\omega_{IR}}{k_{b}T}\int_{0}^{\infty} dte^{-i\omega_{IR}t}R_{pqr}^{(2)}(t)
\end{equation}
\begin{equation}
R_{pqr}^{(2)}(t) = \left\langle A_{pq}(t)M_{r}(0)\right\rangle
\end{equation}
where $n_{BE}(\omega) = 1 - e^{(-\beta\hbar\omega)}$ is the Bose Einstein (BE) factor,\cite{ramirez_quantum_2004} $A_{pq}(t)$ is the $pq$ component of the polarizability of the system at time $t$, $M_{r}(0)$ is the cartesian component $r$ of the total polarization at time 0, $\left\langle\ldots\right\rangle$ is the thermal average, $k_b$ is Boltzmann's constant, $T$ is the respective temperature of the simulation, and $\omega$ represents the frequency of the IR pulse, as the SFG signal is enhanced when $\omega$ is resonant with an interfacial molecular vibration. The polarization and polarizability of the system are the sum of the molecular dipole moments ($\bm{\mu}$) and molecular polarizabilities ($\bm{\alpha}$) respectively:
\begin{equation}
\left\langle A_{qp}(t)M_{r}(0)\right\rangle = \left\langle \sum_i \bm{\alpha}_{i,pq}(t)\sum_j \bm{\mu}_{j,r}(0)\right\rangle
\end{equation}

We carry out a careful calculation of $\chi^{(2)}$ for the slab geometry because contributions from opposite interfaces can interfere destructively, resulting in a vanishing response. Assuming the slab thickness is sufficient to include a bulk region in the middle and a molecular decomposition of $M$ is available, we invert the sign of the molecular dipoles below the center of mass of the slab to prevent this cancellation.\cite{litman_fully_2023}

\subsection*{Numerical Modeling of IR and Raman Spectra}
Within linear response theory, the IR spectrum can be calculated from the Fourier transform of the dipole autocorrelation function:\cite{mcquarrie_statistical_2000, bornhauser_ir}
\begin{equation}
R_{IR}(\omega)n(\omega) \propto \frac{4\pi^2\omega^2}{3k_{B}cnT}\int_{0}^{\infty} dte^{-i\omega t}\left\langle\sum_{i,j}\bm{\mu_i}(0)\bm{\mu_j}(t)\right\rangle
\end{equation}
where $c$ is the speed of light, and $n$ is the refractive index. It is convenient to decompose the polarizability tensor into an isotropic part $\overline{\alpha} = \frac{1}{3} Tr\bm{\alpha}$ and an anisotropic (depolarized) part $\bm{\beta}$ $=$ $\bm{\alpha}$ $-$ I$\overline{\bm{\alpha}}$. Again, within linear response theory, the isotropic and depolarized Raman spectra can be calculated from the Fourier transform of the autocorrelation functions:\cite{msommers_raman_2020}
\begin{equation}
R_{iso}(\omega) \propto \int_{0}^{\infty} dte^{-i\omega t}\left\langle\sum_{i,j}\overline{\bm{\alpha_i}}(0)\overline{\bm{\alpha_j}}(t)\right\rangle
\end{equation}

\begin{equation}
R_{aniso}(\omega) \propto \int_{-\infty}^{\infty} dte^{-i\omega t}\frac{2}{15}Tr\left\langle\sum_{i,j}\bm{\beta_i}(0)\bm{\beta_j}(t)\right\rangle
\end{equation}
\begin{acknowledgement}

This work was supported as part of the Center for Enhanced Nanofluidic Transport (CENT), an Energy Frontier Research Center funded by the U.S. Department of Energy, Office of Science, Basic Energy Sciences under Award DE-SC0019112. Computational support is from the LLNL Grand Challenge Program. The work at the Lawrence Livermore National Laboratory was performed under the auspices of the U.S. Department of Energy under Contract DE-AC52-07NA27344.
The work performed at UC Davis was supported by the National Science Foundation under Grant No. 2305164.

\end{acknowledgement}

\begin{suppinfo}

Snapshots of sampled ice structures, IR and Raman benchmarks using the DNNP@SCAN MLP, bilayer-by-bilayer analysis of IR and Raman of sampled structures, size convergence check, full frequency SFG spectra of sampled ice structures, VDOS assignment of low-frequency region as well as of various facet orientations.  

\end{suppinfo}

\bibliography{sfgbib2}

\end{document}


\renewcommand{\thepage}{S\arabic{page}}  
\renewcommand{\thesection}{S\arabic{section}}   
\renewcommand{\thetable}{S\arabic{table}}   
\renewcommand{\thefigure}{S\arabic{figure}}
\def\theequation{S\arabic{equation}}

\begin{center}
  \begin{figure}[h!]
    \centering
    \includegraphics[width=0.8\linewidth]{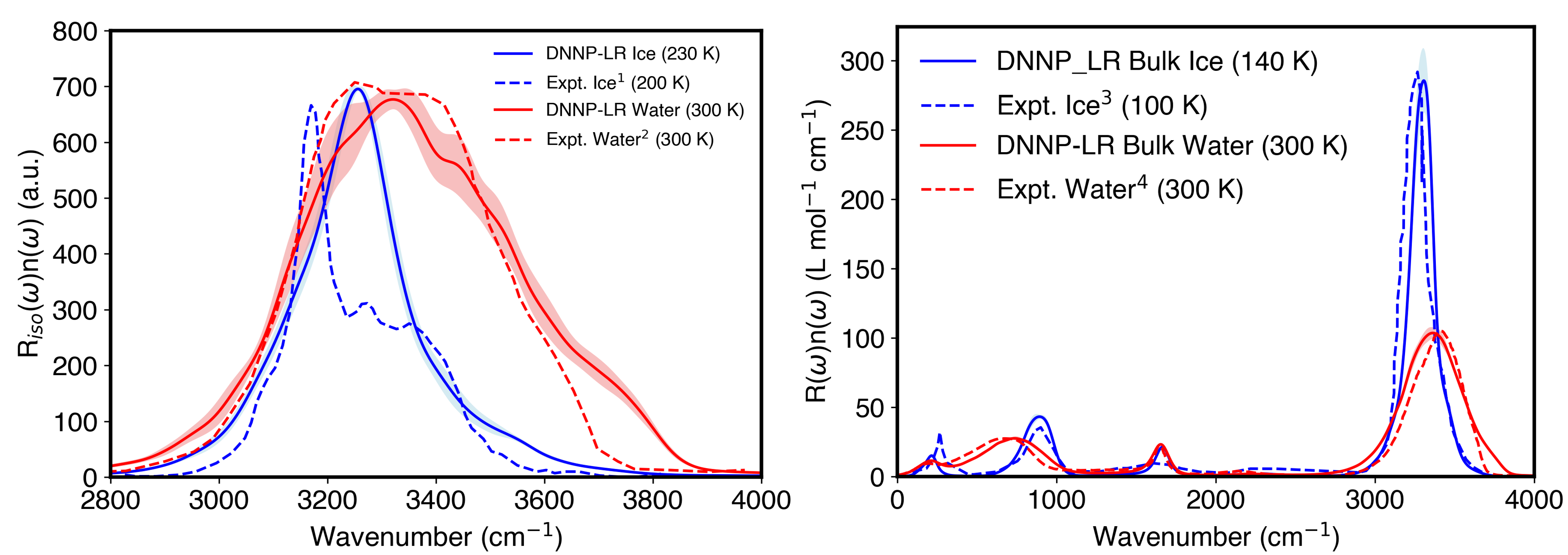}
    \caption{Comparison of the MLP used to experiment for bulk water\cite{bertie_infrared_1996, brooker_raman_1989} and ice\cite{moberg_molecular_2017, bertie_absorptivity_1969} systems for both the Raman and IR spectra. The MLP is within reasonable agreement for both bulk spectra, with slight shifts in peak position that is thought to be attributed to the lack of Nuclear Quantum Effects accounted for in the simulations.}
    \label{fig:bulk_compar}
  \end{figure}
 \end{center}

\begin{center}
  \begin{figure}[h!]
    \centering
    \includegraphics[width=0.8\linewidth]{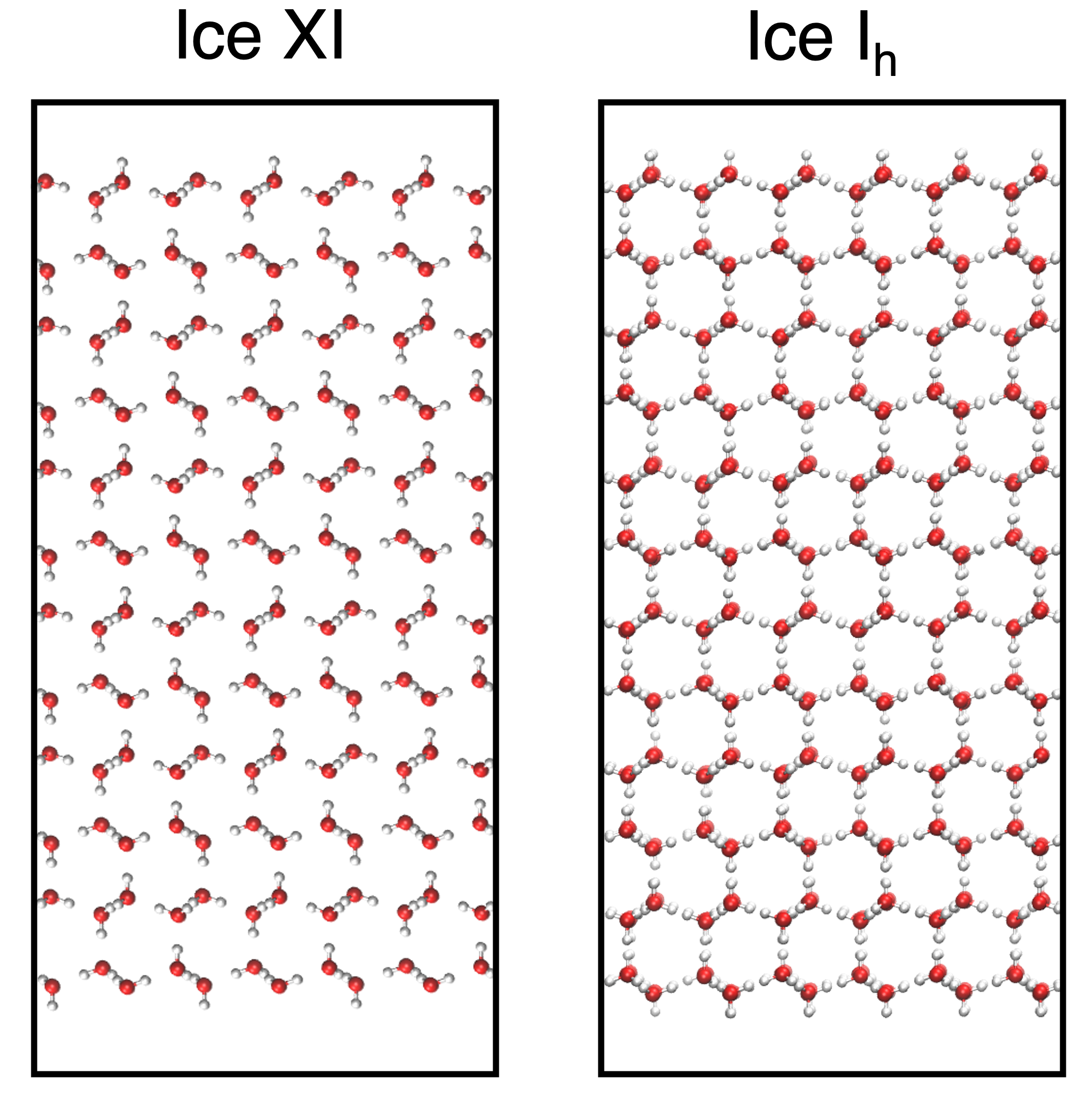}
    \caption{Side-view snapshots of the Ice XI and Ice I$_h$ slabs used in this study.}
    \label{fig:side_view_snap}
  \end{figure}
 \end{center}

\begin{center}
  \begin{figure}[h!]
    \centering
    \includegraphics[width=0.8\linewidth]{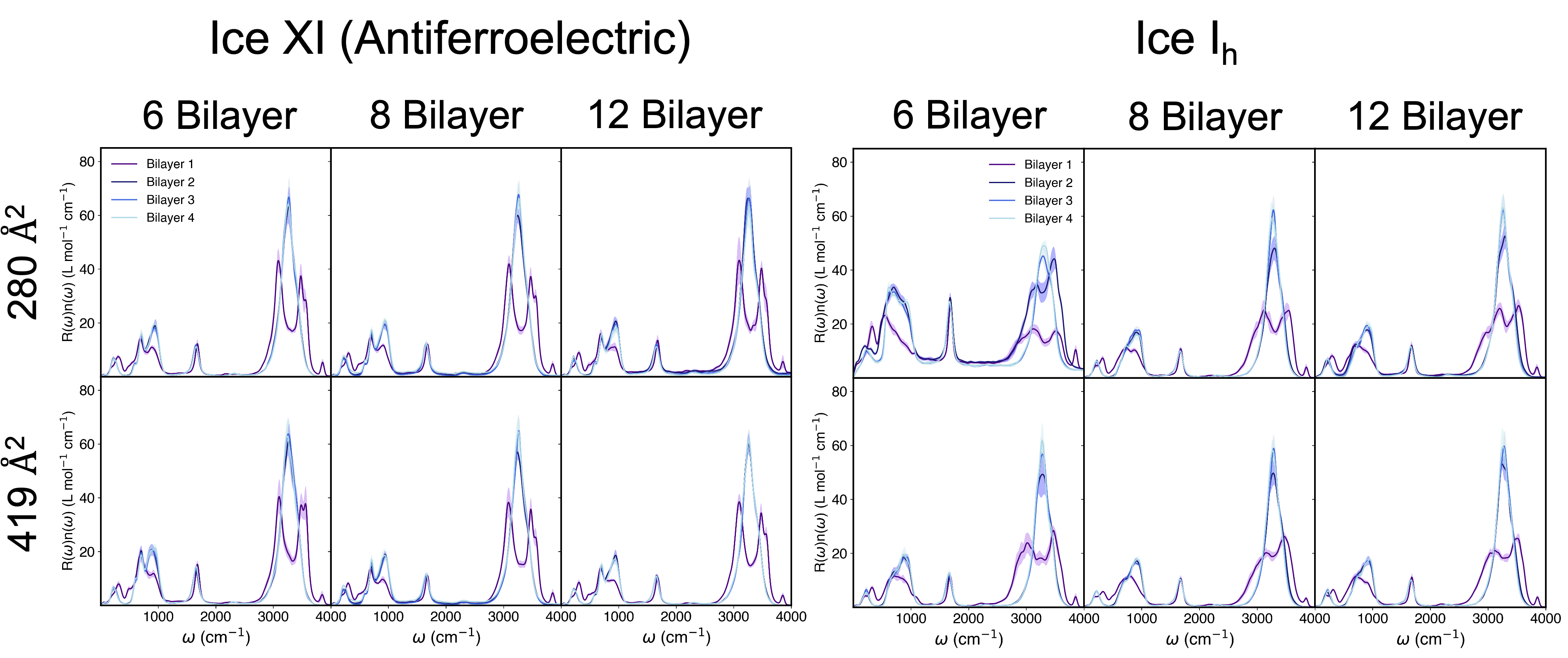}
    \caption{Convergence of the bilayer-by-bilayer IR spectra for both the Ice XI and Ice I$_h$ structures. Systems are converged with the largest sampled surface area and 8 bilayers. For the calculations in this work, the 12 bilayer structures were used. Shaded regions represent the standard deviation over blocks of 30 ps over the total 300 ps.}
    \label{fig:size_converge}
  \end{figure}
 \end{center}

\begin{center}
  \begin{figure}[h!]
    \centering
    \includegraphics[width=0.8\linewidth]{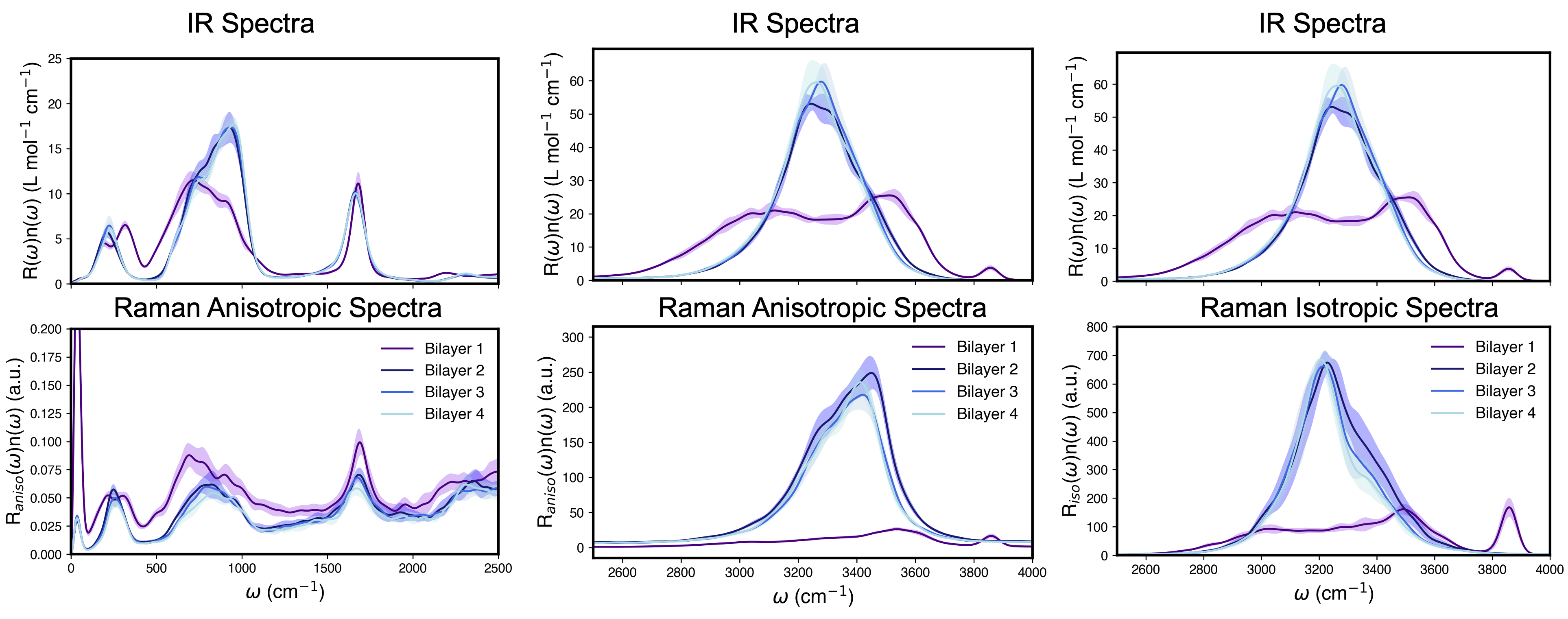}
    \caption{Bilayer-by-bilayer IR, Anisotropic Raman, and Isotropic Raman spectra for an Ice I$_h$ structure. It can be seen that the top bilayer is the most significantly different, while bilayers 2-4 (into the bulk) have bulk vibrational responses. Shaded regions represent the standard deviation over blocks of 30 ps over the total 300 ps.}
     \label{fig:bilayerir}
  \end{figure}
 \end{center}

\begin{center}
  \begin{figure}[h!]
    \centering
    \includegraphics[width=0.8\linewidth]{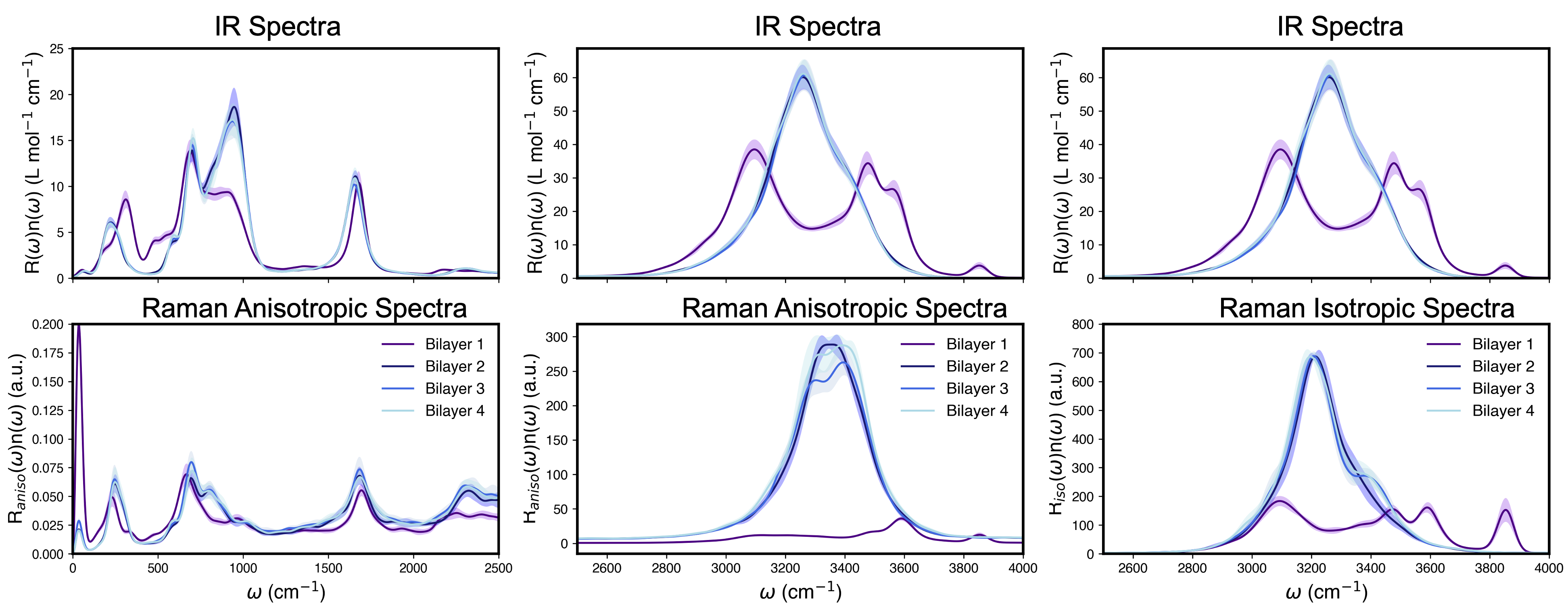}
    \caption{Bilayer-by-bilayer IR, Anisotropic Raman, and Isotropic Raman spectra for an Ice XI structure. It can be seen that the top bilayer is the most significantly different, while bilayers 2-4 (into the bulk) have bulk vibrational responses. Shaded regions represent the standard deviation over blocks of 30 ps over the total 300 ps.}
    \label{fig:icexi_raman_ir}
  \end{figure}
 \end{center}

 \begin{center}
  \begin{figure}[h!]
    \centering
    \includegraphics[width=0.8\linewidth]{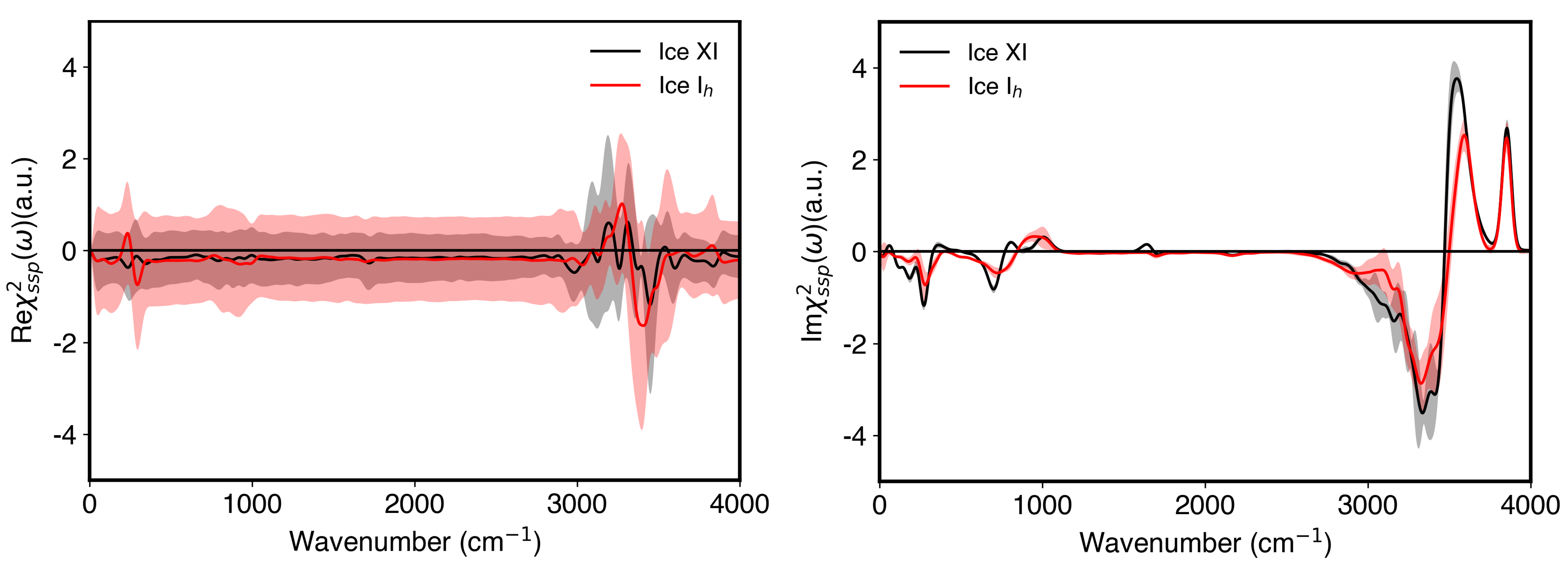}
    \caption{Full frequency (0 - 4000 cm$^{-1}$) Re$\chi_{ssp}^2$ and Im$\chi_{ssp}^{2}$ spectrum of Ice XI and Ice I$_h$. Shaded regions represent averages over the respective sampled simulations.}
    \label{fig:full_spectrum}
  \end{figure}
 \end{center}

\begin{table}[h!]
  \caption{Hydrogen bond distance of the eight identified OH bond types at the basal surface.}
  \label{tbl:hbondlengths}
   \hskip-0.3cm
  \begin{tabular}{lcccc}
    \hline
    \thead{Bond Type} & \thead{VDOS Peak\\ Position (cm$^{-1}$)}  &  \thead{Hydrogen Bond\\ Length ($\AA$)} & \thead{ Cos($\theta$)} \\
    \hline
    \thead{i) Bond 1} & \thead{3860}& \thead{ - } & \thead{0.944}\\
    \thead{ii) Bond 2}& \thead{3586}&\thead{ 1.89 $\pm$ 0.13 }&\thead{ 0.348}\\    
    \thead{iii) Bond 2}   & \thead{3581}  &  \thead{1.88 $\pm$ 0.13} & \thead{-0.271}\\
    \thead{iv) Bond 2} & \thead{3499} &  \thead{1.85 $\pm$ 0.12} & \thead{0.406} \\
    \thead{iii) Bond 1} & \thead{3477} &  \thead{1.83 $\pm$ 0.11} & \thead{-0.209}\\
    \thead{iv) Bond 1} & \thead{3252} &  \thead{1.72 $\pm$ 0.10} & \thead{-0.661}\\
    \thead{i) Bond 2} & \thead{3110} &  \thead{1.67 $\pm$ 0.090} & \thead{-0.345}\\
    \thead{ii) Bond 1} & \thead{3139} &  \thead{1.66 $\pm$ 0.086} & \thead{0.284}\\
    \hline
  \end{tabular}
\end{table}

 \begin{center}
  \begin{figure}[h!]
    \centering
    \includegraphics[width=1.0\linewidth]{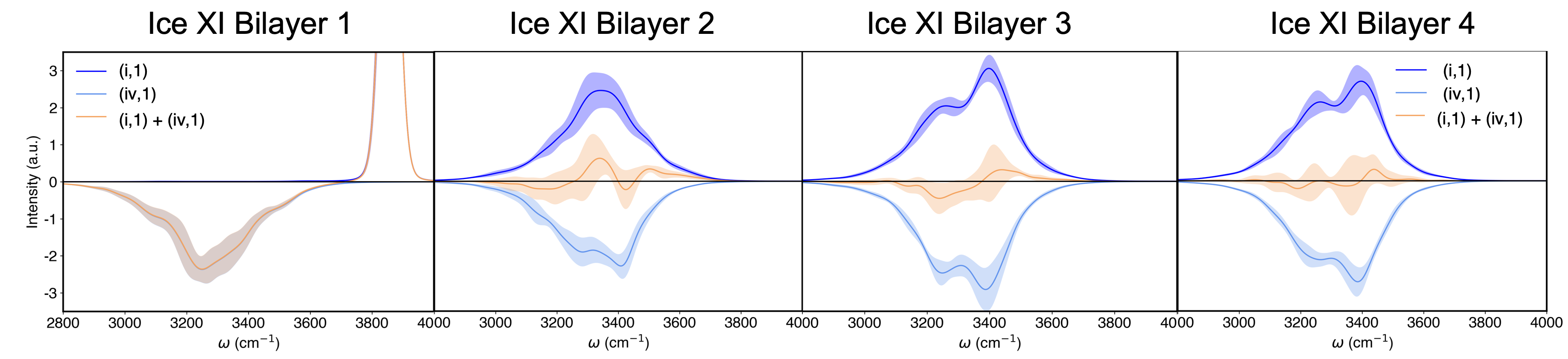}
    \caption{VDOS spectrum of individual OH bonds of the two different water molecule orientations (\romannumeral 1, 1) and (\romannumeral 4, 1) and their sum, for bilayers 1, 2, 3, and 4 of Ice XI. These molecule orientations are the main contributors to the VDOS spectrum for the bulk bilayers (bilayers 2-4), as they represent the molecules that make up the interstitial region between bilayers. Their sum serves as an estimation of the total contributions of each bilayer. It can be seen that the VDOS of bilayers 3 and 4 are qualitatively similar and their sums result in negligible total contributions (as seen in the bilayer analysis of the SFG in Figure 3). While bilayer 2 emphasizes how the interaction between the topmost molecules (\romannumeral 1, 1) of bilayer 2 and the bottommost molecules (\romannumeral 4, 1) of bilayer 1 leads to a lack of centrosymmetry, resulting in a small but non-negligible contribution to the total vibrational spectrum from bilayer 2, that are further enhanced by electronic contributions as seen in the Figure 3. Shaded regions represent the standard deviation over blocks of 30 ps over the total 300 ps.}
    \label{fig:bilayer_contrib_SI}
  \end{figure}
 \end{center}

\begin{center}
  \begin{figure}[h!]
    \centering
    \includegraphics[width=0.8\linewidth]{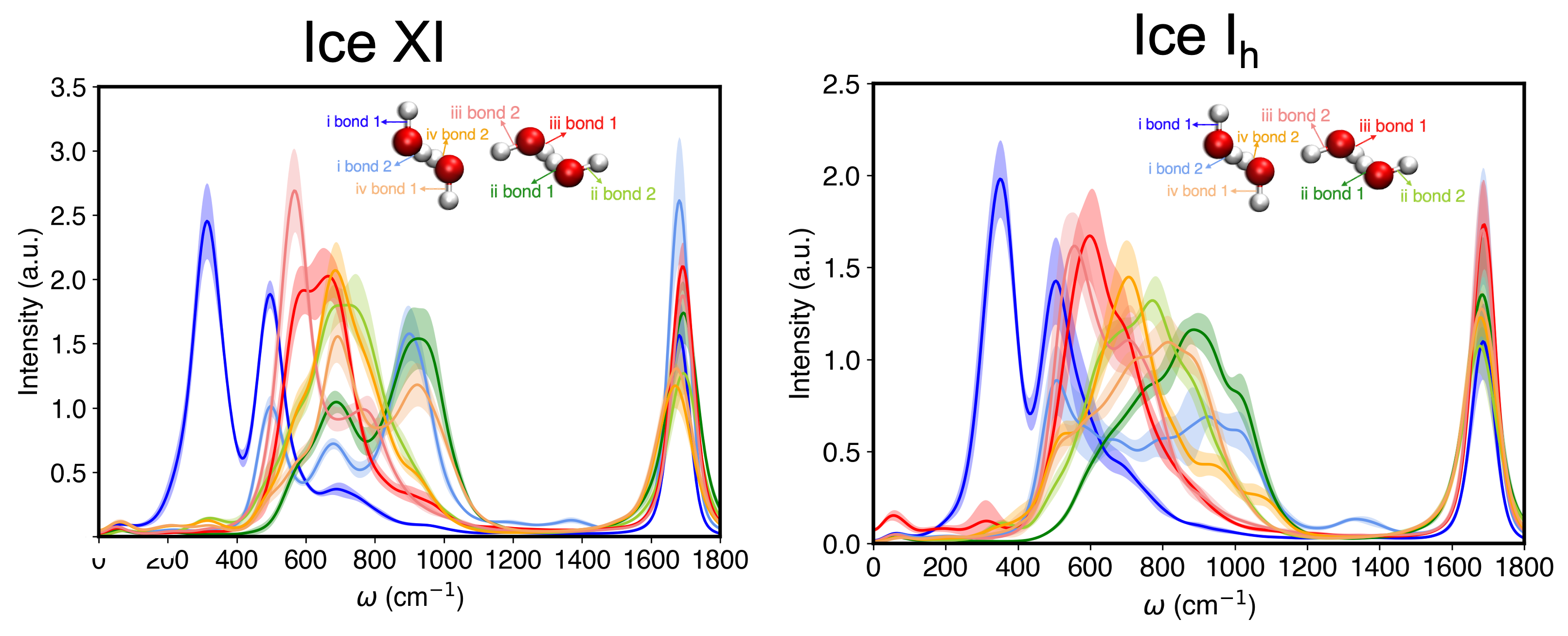}
    \caption{Low frequency region of the VDOS spectrum of individual OH bonds of the four different water molecule orientations at the surface of Ice XI and Ice I$_h$. Shaded regions represent the standard deviation over blocks of 30 ps over the total 300 ps.}
     \label{fig:low_freq_vdos}
  \end{figure}
 \end{center}

\begin{center}
  \begin{figure}[h!]
    \centering
    \includegraphics[width=0.8\linewidth]{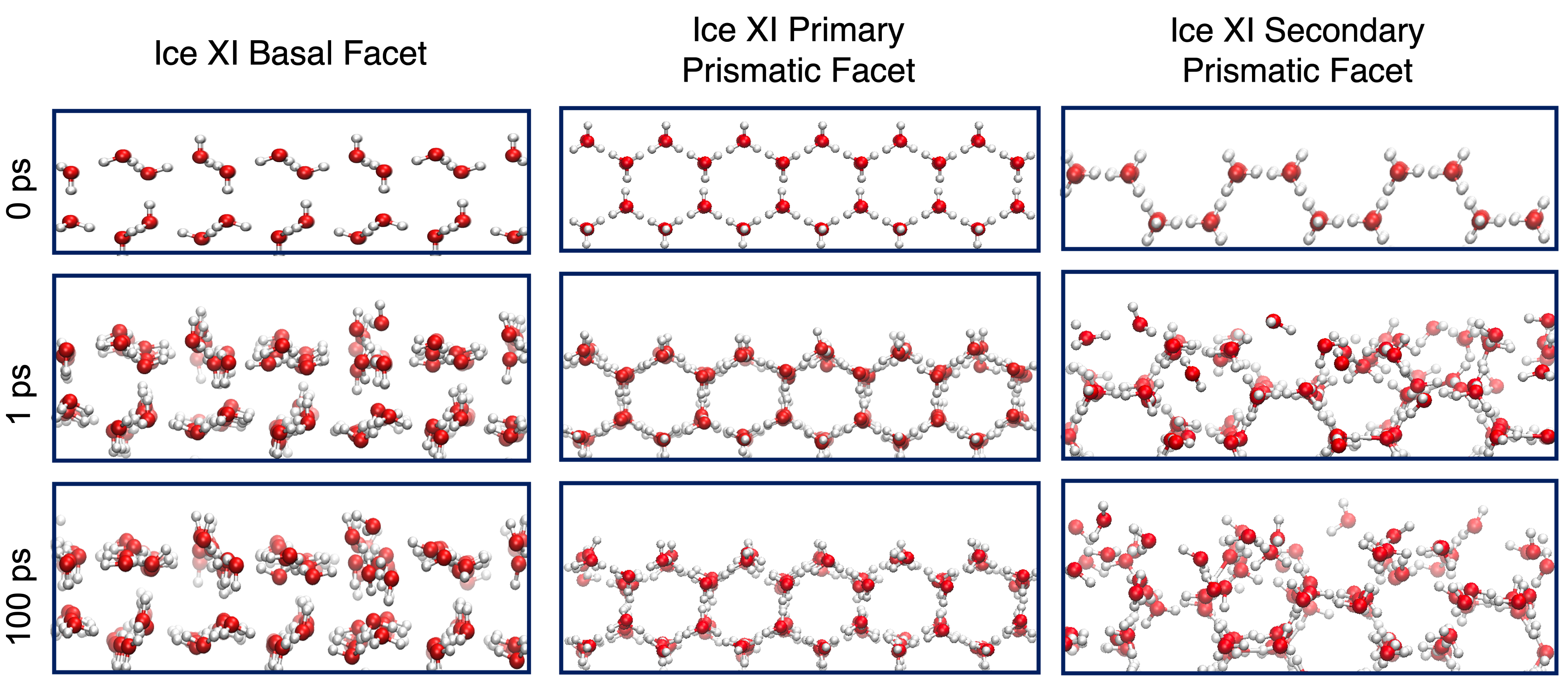}
    \caption{Snapshots of the equilibration runs of the surface of the three sampled low index facets of the Ice XI structures from three time points of a molecular dynamics simulation. At 100 ps, it can be seen that the basal and primary prismatic facet remain crystalline and similar ordering to the starting crystalline structure. While the secondary prismatic facet has some reorganiztion and disorder of the surface most molecules. }
    \label{fig:facet_structure}
  \end{figure}
 \end{center}

 \begin{center}
  \begin{figure}[h!]
    \centering
    \includegraphics[width=0.8\linewidth]{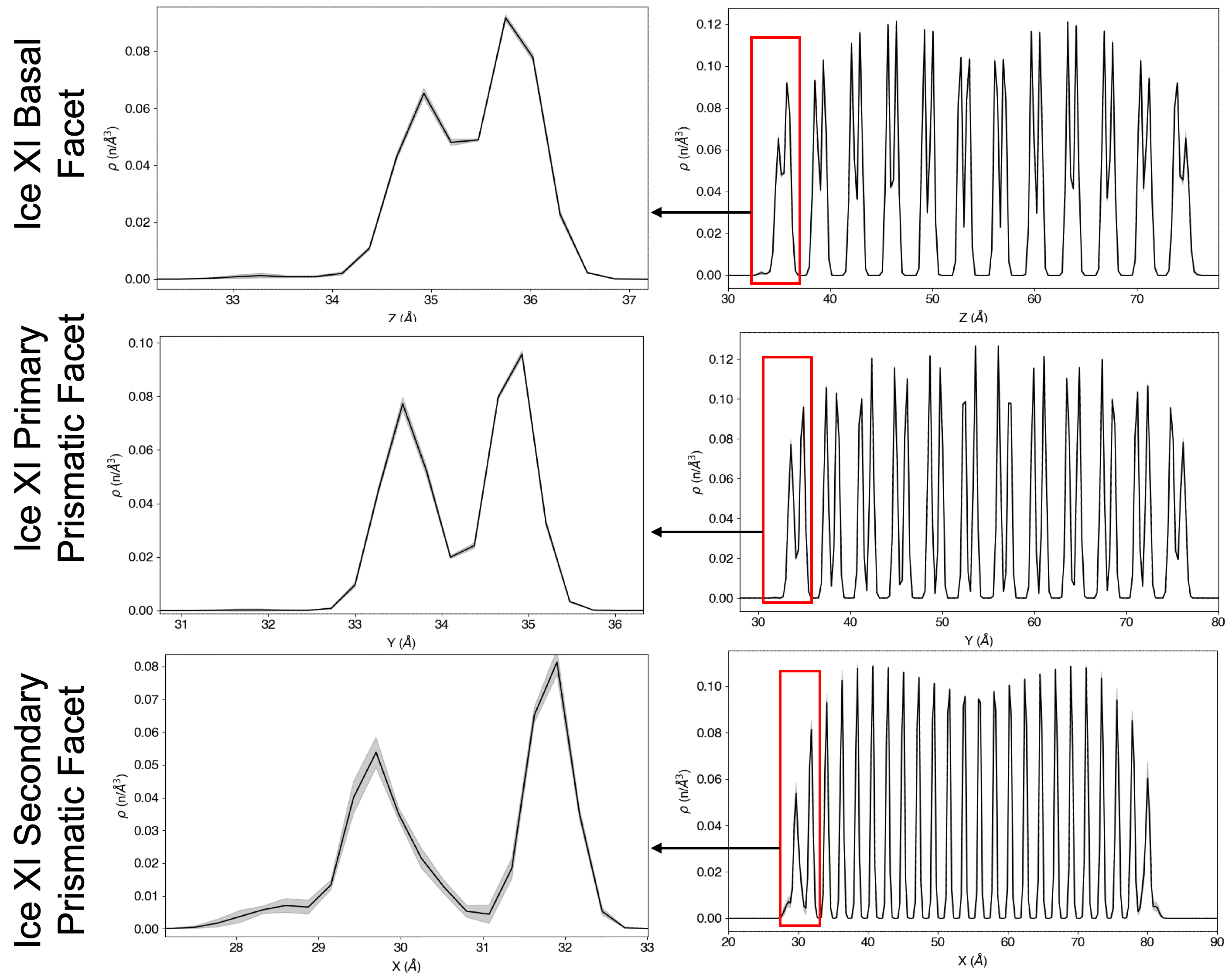}
    \caption{Density profiles for the three low index facets of Ice XI structures from the 300 ps VDOS simulations. Snapshots zoomed in to the surface most layers are included to confirm the more ordered and crystalline structure for the basal and primary prismatic facet  and the restructuring and slight disorder of the oxygen atom lattice and protons for the secondary prismatic facet.}
    \label{fig:densityprofielsice}
  \end{figure}
 \end{center}

  \begin{center}
  \begin{figure}[h!]
    \centering
    \includegraphics[width=0.8\linewidth]{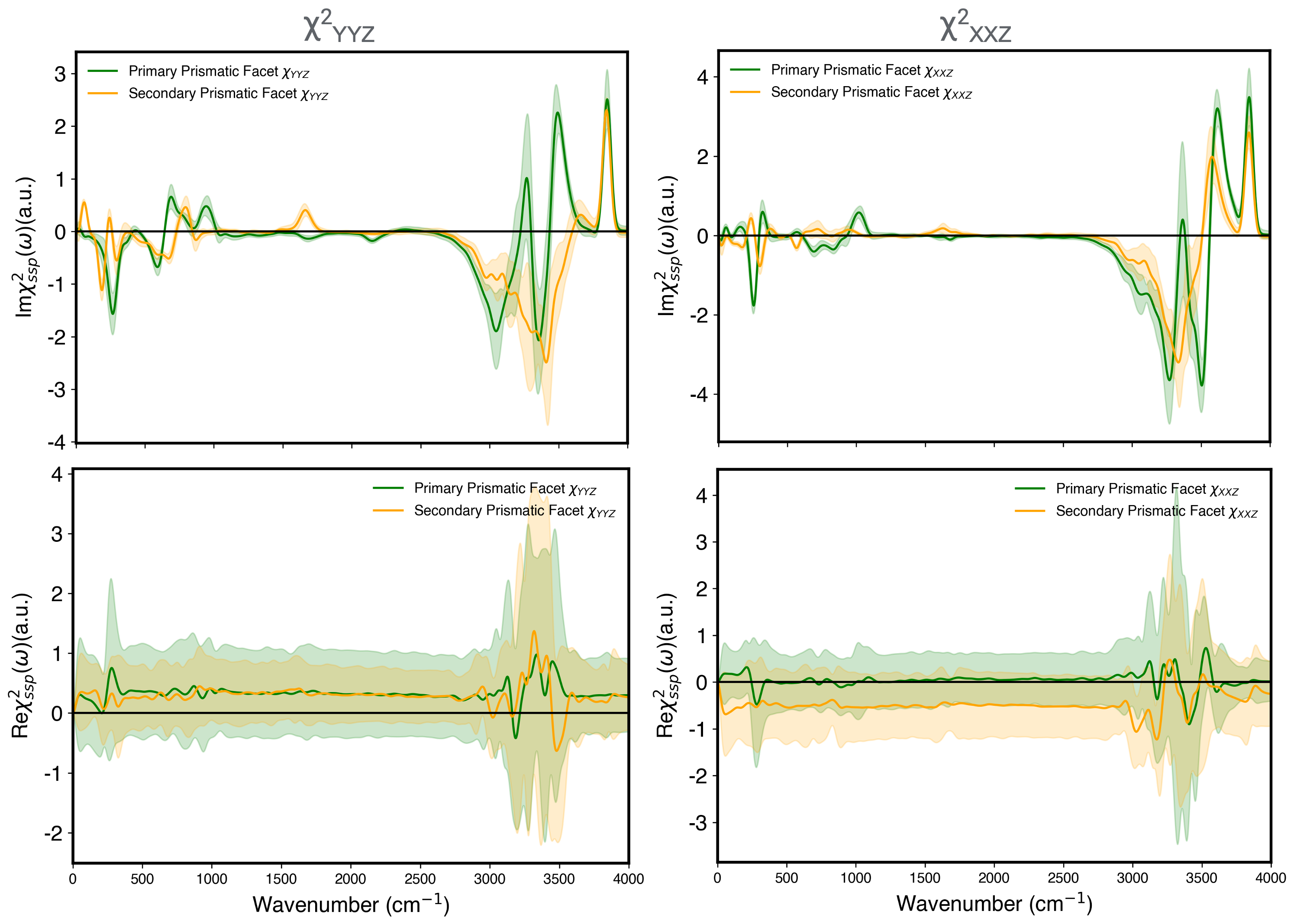}
    \caption{Full frequency (0 - 4000 cm$^{-1}$) Re$\chi_{ssp}^2$ and Im$\chi_{ssp}^{2}$ spectrum of the primary prismatic and secondary prismatic facet of Ice XI, for $xxz$ and $yyz$ polarizations. Shaded regions represent averages over the respective sampled simulations.}
    \label{fig:real_imag_facet_sfg}
  \end{figure}
 \end{center}

  \begin{center}
  \begin{figure}[h!]
    \centering
    \includegraphics[width=0.8\linewidth]{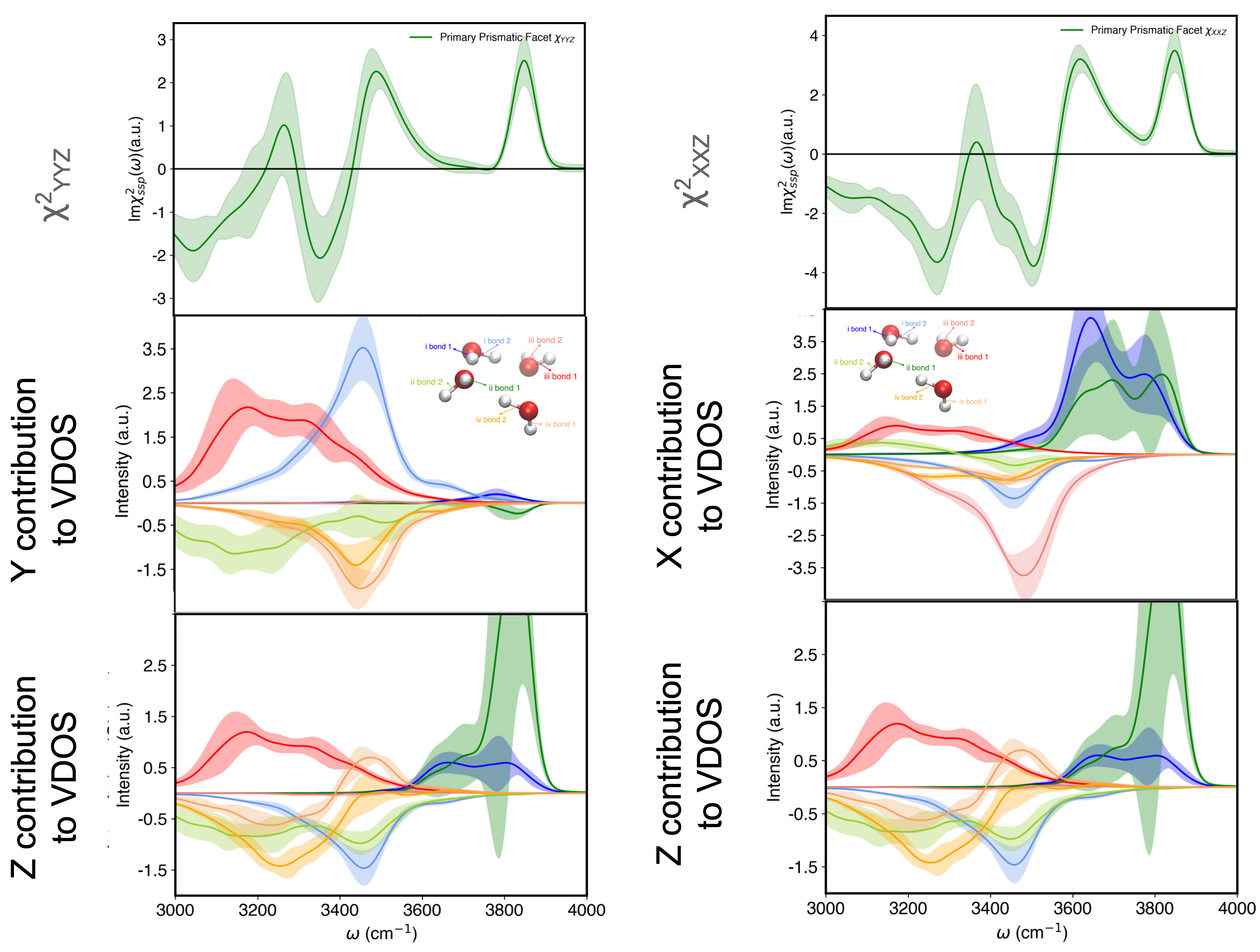}
    \caption{VDOS of individual OH bonds of the four different water molecule orientations at the surface of the primary prismatic facet of Ice XI multiplied by the respective cos($\theta$) of the angle between the individual OH bond and surface normal in the OH stretching region of the vibrational spectra. Shown are the four types of water molecule orientations (top) the VDOS assignment of the OH stretching region. Individual contributions (X, Y, Z) to the VDOS are shown to the relevant polarization of the VSFG. Shaded regions represent the standard deviation over blocks of 30 ps over the total 300 ps.}
    \label{fig:prim_vdos_assign}
  \end{figure}
 \end{center}

\clearpage
\bibliography{sfgbib2}